\documentclass[12pt]{iopart}
\usepackage{epsfig}

\begin{document}

\title{Gas permeation through a polymer network}
\author{B. Schmittmann$^{1}$\footnote{e-mail: schmittm@vt.edu}, 
Manoj Gopalakrishnan$^{2}$, and R. K. P. Zia$^{1}$}

\address{$^{1}$Physics Department and
Center for Stochastic Processes in Science and Engineering,\\
Virginia Tech, Blacksburg, VA 24061-0435, USA\\
$^{2}$MPI f\"{u}r Physik komplexer Systeme, N\"{o}thnitzer Str.
38, 01187 Dresden, Germany}

\begin{abstract}
We study the diffusion of gas molecules through a two-dimensional network of
polymers with the help of Monte Carlo simulations. The polymers are modeled
as \emph{non-interacting} random walks on the bonds of a two-dimensional
square lattice, while the gas particles occupy the lattice cells. When a
particle attempts to jump to a nearest-neighbor empty cell, it has to
overcome an energy barrier which is determined by the number of polymer
segments on the bond separating the two cells. We investigate the gas
current $J$ as a function of the mean segment density $\rho$, the polymer
length $\ell$ and the probability $q^{m}$ for hopping across $m$ segments.
Whereas $J$ decreases monotonically with $\rho$ for fixed $\ell$, its
behavior for fixed $\rho$ and increasing $\ell$ depends strongly on $q$. For
small, non-zero $q$, $J$ appears to increase slowly with $\ell$. In
contrast, for $q=0$, it is dominated by the underlying percolation problem
and can be non-monotonic. We provide heuristic arguments to put these
interesting phenomena into context.
\end{abstract}


\date{January 4, 2004}

\section{Introduction}

Permeation of gas molecules through an amorphous polymer film is a problem
of considerable scientific interest, as well as industrial significance \cite%
{Ref1}. For example, the selective permeability of certain types of
polymeric materials is often used for gas separation. It is therefore of
primary importance to understand how the diffusivity of different gases
depend on various factors, such as the temperature, the size of the
penetrant molecule, and the total amount and distribution of accessible free
volume inside the polymer matrix \cite{Ref2}. In a typical gas permeation
experiment, a polymer film, formed by cooling the polymer from the rubbery
state below the glass transition, is subjected to a pressure gradient across
the film, so that gas molecules permeate the film. After a time lag $\tau $,
the pressure on the low-pressure side starts to increase, and eventually, a
stationary gas current across the membrane is established. For very long
times, the pressures on both sides of the film equilibrate; this late stage
is not of interest here. The important parameters which are often used 
to describe the permation process are (\emph{i}) the solubility $S$,
which gives the total amount of gas trapped in the membrane and (\emph{ii})
the diffusion coefficient $D$ of a gas molecule inside the network. To
determine the latter, one approximates the late-time dependence of the
pressure by a straight line and obtains its intercept, $t_{o}$, with the
zero pressure axis. The diffusion coefficient then follows from the relation 
$t_{o}=L^{2}/6D$, where $L$ is the thickness of the film. The product of the
solubility and diffusion coefficient is referred to as the permeability $%
P=DS $, and is a direct measure of the efficiency of the permeation process.

From a slightly different point of view, the gas permeation problem may be
discussed in the language of (electric)\ resistor networks. The pressure
gradient set up across the membrane is analogous to a fixed voltage
difference. Local configurational fluctuations inside the polymer matrix
allow gas molecules to jump across energy barriers between voids, forming a
connected set of ``resistors'' which control the flow of the gas current. In
fact, passage of (electric)\ current through a set of \emph{randomly }%
distributed resistors is a problem that has been extensively studied in
statistical physics \cite{Ref3}. From this perspective, gas diffusion
through a polymer network is quite intriguing since the matrix of energy
barriers and voids can hardly be considered random or uncorrelated.

It is therefore natural to explore the role of these correlations in the
transport process. The primary objective of this paper is to introduce a
lattice-gas model with correlated, quenched energy barriers to study the
transport of gas molecules through the polymer network. While we simplify
the description significantly, by reducing it to two dimensions, and
modeling polymers as simple non-interacting random walks, we nevertheless
retain an essential feature of the original problem, namely, the effect of
correlations induced by the polymer connectivity. We find that the currents
through this network depend very sensitively on (\emph{i}) whether particle
motion can be completely blocked by certain energy barriers; and (\emph{ii})
whether these spatially correlated barriers can percolate across the
lattice. Our discussion draws on an earlier study \cite{Ref4} in which we
investigated the percolation properties of random walks of varying length.

The remainder of this paper is divided into three sections: Section II
contains the model description, Section III discusses the Monte Carlo
simulation results and Section IV outlines our conclusions.

\section{The model}

We consider a polymer network whose large scale structure is static over the
time scale of the simulations. This is a reasonable simplification since
typical permeation experiments use polymers in the glassy state: Such
membranes offer better selectivity than those in a rubbery state. Small 
\emph{local} rearrangements -- which let gas molecules pass from one void to
another -- do occur and will be modeled via a distribution of energy
barriers which the molecules need to overcome. The polymer network is
generated by placing $N$ non-interacting random walks of length $\ell $
(referred to as ``$\ell $-mers'') on a two-dimensional square lattice with $%
L^{2}$ sites. Each \emph{bond }of the lattice can carry one or more segments
(i.e., steps of the random walk), with an average segment density per bond
of $\rho =N\ell /2L^{2}$. The gas molecules are modeled as hard-core
particles which occupy the\emph{\ cells }of the lattice, and no more than
one particle is allowed per cell.

While modeling polymers as non-interacting (rather than self-avoiding) walks
is almost certainly an oversimplification, we can offer two motivations for
such a simple description. First, real membranes are produced by rapid
cooling of a rubbery melt; hence, typical polymer configurations are
``frozen'' high-temperature states for which interactions are less
important. Second, when discretizing a real network, it is natural to choose
the lattice spacing to be comparable to the polymer persistence length. In 
this case, a polymer ``segment'' contains many monomers. Since the persistence
length is typically much larger than the size of a monomer unit, we have 
in effect adopted a coarse-grained description. On this length scale, 
self-avoidance and other complications may be neglected.

Each polymer segment constitutes an energy barrier in the path of a gas
molecule. The dynamics of the molecules is modeled as activated hopping over
these barriers. Once a particle-hole pair, separated by such a barrier, is
chosen, the probability for them to exchange is $q^{m}$, where $m$ is the
number of segments in the barrier and $q$ is just $e^{-\beta \epsilon }$,
with $\epsilon $ being an energy scale and $\beta =1/k_{B}T$ being the
inverse temperature. As an alternate view, one may consider the particle
motion as resulting from the short time scale segmental motion of the
polymer. If the probability of displacement of a single segment is $q$, then 
$q^{m}$ would be the probability of simultaneously displacing all $m$
segments.

To model the pressure gradient, the cells in the first row ($y=0$) of the
lattice are kept filled at all times. At the beginning of the simulation
(time $t=0$), the remainder of the lattice is empty. At times $t>0$, each
particle randomly selects a direction of motion with probability $1/4$, and
attempts to cross the barrier between its originating and target cells.
Provided the target cell is empty, the particle will be moved with
probability $q^{m}$. After every particle in the system has attempted a move
(on average), $t$ is incremented by one Monte Carlo step (MCS). When a
particle arrives in a cell within the last row ($y=L$), it is removed
instantaneously from the lattice. Clearly, this sets up a density gradient
across the system which drives a particle current in the $y$-direction.

In the simulations, we measure the following quantities:

\begin{enumerate}
\item The number of particles which have entered the system up to time $t$,
denoted by $M_1(t)$

\item The number of particles which have exited the system up to time $t$,
denoted by $M_2(t)$.

\item The mean first passage time, $\tau $, for the arrival of the \emph{%
first} particle at the far boundary $y=L$.
\end{enumerate}

At any time $t$, therefore, the number of particles absorbed by the polymer
network is given by $M(t)=M_{1}(t)-M_{2}(t)$. At sufficiently late times,
the system approaches a steady state, which is characterized by a
time-independent solubility 
\begin{equation}
M(\rho ,\ell ;q)\equiv \lim_{t\rightarrow \infty }M(t)
\end{equation}%
and a constant particle current, or permeability, $J(\rho ,\ell ;q)$,
defined via 
\begin{equation}
J(\rho ,\ell ;q)\equiv \lim_{t\rightarrow \infty }\partial _{t}M_{2}(t)
\end{equation}%
To give a reference point, the current through an empty lattice would be $%
1/2 $. The mean first-passage time provides information about the typical
path length travelled by a particle, and how quickly it hops over the
barriers encountered along that path.

The numerical simulations were carried out for various values of the segment
density $\rho $ and polymer length $\ell $. We choose a lattice with $%
128\times 128$ sites for the simulations, which was also the largest lattice
simulated in \cite{Ref4}. We choose polymer lengths in powers of $4$: $\ell
=4^{n}$ with $n=0,1,2,3,4$. The typical span of even the longest polymer ($%
\ell =256$) is much smaller than the lattice dimension, so that a single
polymer only very rarely extends across the whole lattice. Our choices for
the segment density $\rho $ were guided by the phase diagram of the $q=0$
system which we will summarize in the next section. The smallest value of $%
\rho $ was $0.5$ and the largest was $1.0$. Intermediate values (e.g. $\rho
=0.71$) are controlled by $N$, the (integer)\ number of polymers of length $%
\ell $ placed in the system. For different $\ell $'s, we selected $N$'s to
be within less than a percent of $2\rho L^{2}/\ell $. So far, we considered
just two values of the bond-crossing probability: $q=0.0$ and $q=0.1$. Due
to computational limitations, we generated between $15$ and $48$ different
polymer configurations for each set of parameters $\left( \rho ,\ell
,q\right) $. $M(t)$ and $M_{2}(t)$ are collected for $2\times 10^{6}$ MCS
for each configuration and then averaged. 

Let us comment briefly on statistical errors, which can be estimated by 
comparing currents for different configurations at the same 
$\left( \rho ,\ell,q\right) $.
They are smallest (below $5\%$) for the shortest polymers and $q=0.1$. 
As $ \ell$ increases, the errors also become larger, reaching about $20\%$ 
for $ \ell=256$ at this $q$. 
For $q=0.0$ and especially $\rho=0.71$, finite-size effects allow for 
both blocked and open configurations at the same 
$\left( \rho ,\ell,q\right) $, making it 
difficult to assign error bars reliably. A conservative estimate would give  
errors as large as the currents themselves, so that our conclusions 
can only be preliminary until better statistics are available.

\section{Numerical Results}

\subsection{The $q=0$ case: Percolation of occupied bonds}

When the bond-crossing probability $q$ vanishes (or, equivalently, the zero
temperature limit), bonds carrying one or more segments become completely
impenetrable. As a consequence, the gas current must vanish if the occupied
bonds form a connected path (a ``spanning cluster'') transverse to the
density gradient. The problem, therefore, reduces to bond percolation of $N$ 
$\ell $-mers on a square lattice. We studied this problem in \cite{Ref4},
and summarize only the salient results here.

With the segment density (per bond) given by $\rho =N\ell /(2L^{2})$, the
fraction of occupied bonds, $p(\rho ,\ell )$, approaches a well-defined
value in the thermodynamic limit. In our simulations, we found that $p$
decreases with $\ell $ for fixed $\rho $ and increases with $\rho $ for
fixed $\ell $. While the latter behavior is obvious, the former needs some
explanation. To make it more intuitive, let us consider two situations where
the density $\rho $ is kept fixed, but the polymer lengths (and hence, their
numbers) are different. Configurations with few, long polymers are likely to
exhibit numerous bonds which are multiply occupied since a simple random
walk retraces any given step with probability $1/4$. In contrast, such
multiple bond crossings are far less likely for short polymers, even at the
same $\rho $, resulting in a larger $p$. To rephrase, single segments leave
far fewer bonds unoccupied (``free'') than longer polymers.

With a qualitative understanding of $p(\rho ,\ell )$, it is natural to ask
how the percolation threshold, $p_{c}(\ell )$, is crossed as one increases $%
\rho $ at given $\ell $. From simulations, we find that $p_{c}(\ell )$ is a 
\emph{monotonically decreasing} function of $\ell $, i.e., when $\ell $ is
large, a smaller number of occupied bonds already suffices to ensure the
presence of a spanning cluster. We believe that this is primarily due to
long polymers forming extended cigar-shaped objects \cite{Ref5} which
percolate more easily \cite{Ref6} than single occupied bonds or short
polymers.

Clearly, two consequences of polymerization compete with one another:\ the
elongated shape of the polymers tends to lower, while the presence of
multiple occupancies tends to increase, the percolation threshold. Due to
this competition, the critical segment density per bond,$\;\rho _{c}(\ell )$%
, is a \emph{non-monotonic} function of $\ell $, shown in Fig.~1 for a $%
128\times 128$ square lattice. We see that $\rho _{c}(\ell )$
increases from $\ln 2=0.692$ for $\ell =1$ to $0.744$ for $\ell =4$ but
decreases beyond that to $0.724$ ($\ell =16$) and $0.703$ for $\ell =64$.
For shorter polymers, $\rho _{c}(\ell )$ is dominated by the effect of
multiple occupancies (which increase with $\ell $), but for longer ones,
their tendency to form extended objects which percolate more easily, takes
over. When $\ell $ becomes comparable to $L$, finite-size effects become
noticeable, causing $\rho _{c}(\ell )$ to increase again: $\rho
_{c}(256)=0.718$ for $L=128$. Here, multiple occupancies dominate again,
this time due to large sections of \emph{different} polymers overlapping one
another. Not surprisingly, this effect is more pronounced on smaller
lattices, making it difficult to extrapolate the behavior of $\rho _{c}(\ell
)$. Whatever the eventual resolution of these finite-size effects, in the
thermodynamic limit $\rho _{c}(\ell )$ plays the role of a phase boundary: a
gas current can flow only if $\rho <$ $\rho _{c}(\ell )$.

\begin{figure*}[!t]
\begin{center}
\epsfig{file=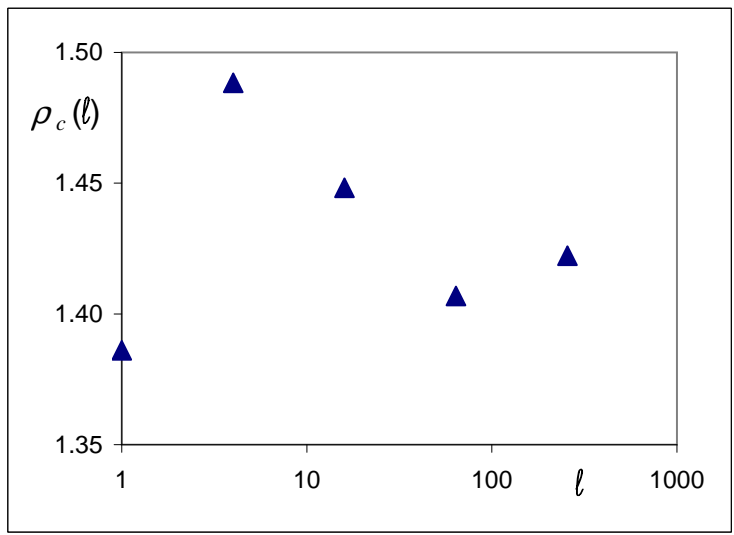, width=4in}
\end{center}
\caption{The critical segment density $\rho_{c}(\ell)$ vs. 
polymer length $\ell$ at the
percolation threshold for system size $128 \times 128$ 
\cite{Ref4}.}
\label{fig1}
\end{figure*}

\subsection{The gas current $J$}

Our earlier findings \cite{Ref4} provide an answer to a simple ``yes/no''
question: Will a gas current flow or not? We now turn to a much more
detailed question: If a current does flow, what are its quantitative
characteristics? By measuring currents, we are probing not only the mere
existence of open channels, but also their number and their length: How
many distinct paths are there for a gas molecule to travel from one edge of
the system to the other, and how many cells will it visit along a given
path?\ Again, we use a combination of simulations and simple intuitive
arguments to explore the answers. We begin with systems in the low
temperature limit $q=0.0$, i.e., occupied bonds cannot be crossed at all,
and then consider the effects of softening this constraint by having
non-zero $q$.

Our numerical results for the steady-state particle current $J(\rho ,\ell
;q) $ are presented in Figs.~2 and 3. In Fig.~2, we show $J$ plotted vs.
polymer length $\ell $, for fixed $\rho =0.5$ and $q=0.0$. According to the
results summarized in the previous section, this value of $\rho $ is well
below the percolation density $\rho _{c}(\ell )$ for all $\ell $ considered
here; hence the current is non-zero for these $\ell $. In contrast, the
inset shows $J$ vs. $\ell $ at a higher density, $\rho =0.71$, which is
within a few percent of the critical $\rho _{c}(\ell )$. In fact, for $\ell
=1$, $\rho $ lies slightly \emph{above} the critical $\rho _{c}(1)=\ln 2$,
and consequently, no current should flow for this $\ell $. As one might
anticipate, the absolute values for the currents are much smaller (by at
least a factor of $10$) for the larger $\rho $, since more bonds are blocked
in that case. On a more quantitative level, we observe very different
behaviors at the two densities. For $\rho =0.5$, $J$ decreases with $\ell $,
for all but the smallest $\ell $. In contrast, at $\rho =0.71$, the current
first increases, then drops off again to a minimum at $\ell =64$, after
which it recovers. Thus, polymerization has very different effects on the
current, at low and high segment density.

\begin{figure*}[!t]
\begin{center}
\epsfig{file=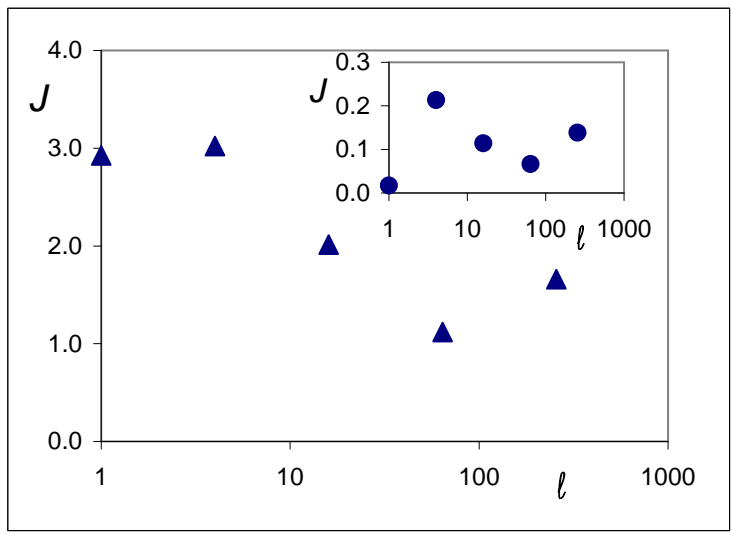, width=4in}
\end{center}
\caption{The gas current $J(\rho,\ell;q)$, integrated over $100$ MCS,
vs. polymer length $\ell$
for $\rho=0.5$ and $\rho=0.71$ (inset), at $q=0.0$ 
for an $128 \times 128$ system. }
\label{fig2}
\end{figure*}

We now attempt to interpret these findings, in the light of our percolation
study summarized above. Considering, first, $\rho =0.5$, we note that this
value is well below $\rho _{c}(\ell )$ for all $\ell $ considered. In other
words, a typical configuration exhibits only finite clusters of occupied
bonds, allowing for numerous current-carrying channels with a length of $%
O(L) $. Polymerizing single segments ($\ell =1$) into short polymers (say, $%
\ell =4$) frees some occupied bonds (by increasing multiple occupancies) but
has little effect on the number or the length of the conducting channels.
Thus, the current for $\ell =4$ is, at best, slightly enhanced over the
current at $\ell =1$. However, as the polymers become longer, they also
extend further, forcing particles to flow around them. Moreover, longer
polymers begin to form ``loops'' which enclose a fair fraction of free, but
now inaccessible, bonds. Hence, having a few long polymers in the system can
increase the effective length of the conducting channels significantly and
thus decrease the current, as illustrated by Fig.~2.

Turning to the inset of Fig.~2, the picture changes drastically. This
density ($\rho =0.71$) is within a few percent of the percolation threshold, 
$\rho _{c}(\ell )$; if any open channels exist, they are rare and fractal.
Even minor rearrangements of occupied bonds suffice to open a new channel,
or block an existing one. Turning to specific $\ell $'s, we recall that $%
\rho =0.71$ is slightly above the percolation threshold for \emph{single}
segments: $\rho _{c}(1)=0.692$. Hence, for $\ell =1$ no current should
flow, at least in the thermodynamic limit. However, finite-size effects wash
out a sharp percolation threshold, so that a very small current is
observed: a few of the simulated polymer configurations did not percolate.

If single segments are tied together into longer polymers with, say $\ell =4$%
, the number of occupied bonds is lowered significantly, since higher
occupancies will be generated. As a result, a much larger segment density is
required for percolation:\ $\rho _{c}(4)=0.744$. Translated into currents,
this implies that the $\ell =4$ system will carry a significant current at $%
\rho =0.71$, as borne out by the data shown in the inset. Yet, the tendency
of polymers to form extended objects competes with this trend. We believe
that the minimum, observed at $\ell =64$, is genuine (even though it lies
well within the error bars) due to the fact that this system is extremely
close to the percolation threshold: $\rho _{c}(64)=0.703$. If it were not
for finite-size effects, we would expect the current to vanish here. To
summarize, $J(0.71,\ell ;0.0)$ is dominated by the behavior of $\rho
_{c}(\ell )$: $J$ vanishes if $\rho >\rho _{c}(\ell )$, and otherwise
roughly follows $\rho _{c}(\ell )$: the distance to the percolation
threshold $\rho _{c}(\ell )$ controls the number of open channels which, in
turn, control the current. Of course, there is no need to simulate systems
with densities $\rho >\rho _{c}(\ell )$ since the currents are expected to
vanish.

Let us now compare these findings to currents measured in systems with
finite energy barriers. Specifically, we choose $q=0.1$ for the rate with
which a singly occupied bond is crossed by a gas molecule. For a bond
occupied by $m$ segments, this rate drops to $q^{m}$. Fig.~3 shows the
corresponding currents for $\rho =0.5$ and $\rho =1.0$ (inset). Both data
sets show currents $J(\rho ,\ell ;0.1)$ which appear to increase gently with 
$\ell $. In terms of absolute values, $J(0.5,\ell ;0.1)$ is larger than $%
J(1.0,\ell ;0.1)$ because fewer bonds are occupied at the lower density. Of
course, $J(0.5,\ell ;0.1)$ is also larger than $J(0.5,\ell ;0.0)$: even
though both cases have (on average) the same number of occupied bonds, these
are blocked completely in the latter, but not the former case.

\begin{figure*}[!t]
\begin{center}
\epsfig{file=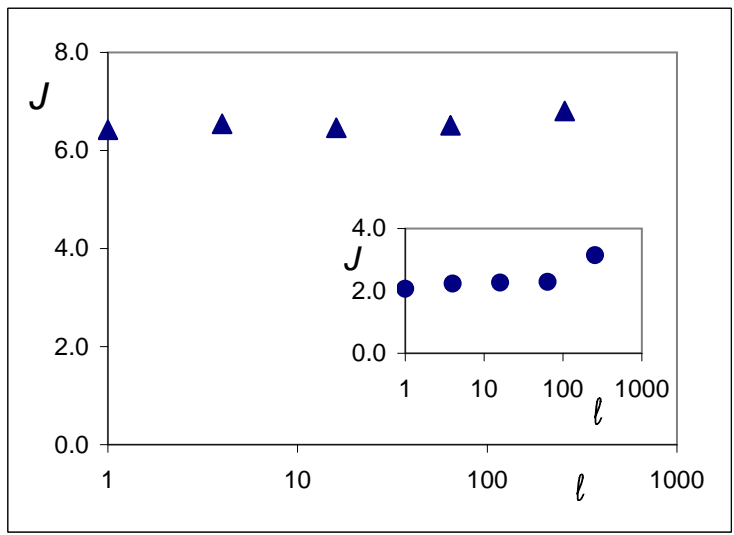, width=4in}
\end{center}
\caption{The gas current $J(\rho,\ell;q)$, integrated over $100$ MCS,
vs. polymer length $\ell$
for $\rho=0.5$ and $\rho=1.0$ (inset), at $q=0.1$
for an $128 \times 128$ system.}
\label{fig3}
\end{figure*}

For $q=0.1$, a singly occupied bond allows gas molecules to cross in one out
of $10$ attempts. In order to block efficiently, a bond should carry, say,
three or more segments, so that the rate of crossing is reduced to $10^{-3}$
or less. While such ``highly occupied''\ bonds exist, of course, they are
extremely unlikely to form large connected clusters, even for the largest $%
\ell $: they constitute less than about $10\%$ of bonds for $\rho =0.5$, and
less than about $15\%$ for $\rho =1.0$. As a result, there are no extended
objects that might impede the flow of particles; instead, the currents are
dominated by the number of completely free ($m=0$) and singly occupied ($m=1$%
) bonds. Unlike the $q=0.0$ case, domains of unoccupied bonds inside polymer
loops can now be accessed and contribute to the currents. This effect is
particularly stark for systems above the percolation threshold: e.g., at
density $\rho =1.0$ with $q=0$, all currents would vanish.

Regarding the small dip observed in the $\rho =0.5$ data, we have no convincing
explanation to offer at this stage. With statistical errors of about $10\%$,
it seems hardly significant. Yet, it is conceivable that the $J(\rho ,\ell
;q)$ data reflect, in some very loose sense, the ``sum'' of a residual
current flowing only across connected free bonds (given by $J(\rho ,\ell
;0.0)$), plus a contribution carried by the remaining (isolated) free bonds
and singly occupied ones. Since the latter is likely to increase
monotonically with $\ell $, while $J(0.5,\ell ;0.0)$ is monotonically
decreasing, one might interpret the dip in $J(0.5,\ell ;0.1)$ as being the
result of ``adding'' the two trends. Better data are needed
before these questions can be settled satisfactorily.

\subsection{Solubility $M$ and first passage time $\protect\tau $}

In very general terms, the current $J$ may be thought of as dependent on two
major factors: how many particles are present inside the matrix at any given
time, and how fast they move across barriers, on average. In order to
quantify these, we measured the particle `solubility' $M(\rho ,\ell ;q)$ in
the steady state and the mean `first passage time' $\tau $, which is the
time it takes for the first particle to emerge out of the network at its far
end. If we adopt a simplistic view and imagine that the (barrier-hopping)
motion of a single particle inside the polymer network is simply diffusion
with an effective diffusion coefficient $D$, then the first passage time $%
\tau $ provides a measure of $D$: longer time intervals $\tau $ imply lower $%
D$.

We find that the solubility $M(\rho ,\ell ;q)$ roughly trails the current
data: Higher currents are associated with larger $M(\rho ,\ell ;q)$, and
vice versa. Given that higher currents imply more open channels (which can
then be filled with additional particles) and fewer (if any)\ inaccessible
domains, this seems plausible. Our data for the mean first passage time in $%
q=0.0$ systems are unfortunately very noisy, so that no reliable conclusions
can be drawn. For the $q=0.1$ systems, $\tau $ decreases with $\ell $,
indicating that particles encounter fewer obstacles on their way through the
network. Again, this broadly mirrors the current data.

\section{Conclusions}

In this paper, we studied a simple model of gas transport through a polymer
membrane. The membrane is modeled as a network of non-interacting random
walks, placed on the bonds of a square lattice in two dimensions. The
network is characterized by the total segment density, $\rho $, and the
length, $\ell $, of the random walks. The gas molecules are modeled as
hard-core particles performing activated hopping from one lattice cell to a
nearest neighbor. When they cross a bond carrying $m$ segments, they
encounter an energy barrier $q^{m}$, with $q=e^{-\beta \epsilon }$. At one
end of the lattice, the particle density is held at unity; at the other end,
particles are immediately removed. These boundary conditions drive particles
across the network. To quantify the transport properties of the network, we
focus primarily on the gas current, $J(\rho ,\ell ;q)$, in the steady state.
We explore in particular how polymerization at fixed segment density (i.e.,
joining a given number of segments into polymers of varying length) affects
the currents. We also measure the total number of particles stored in the
network, as well as the mean first passage time for the first particle to
arrive at the far end of the lattice.

It is immediately obvious that this problem will be most interesting, and
most difficult, in the low temperature regime, when the polymers form
significant obstacles. Considering therefore small values of $q$, including $%
q=0$, our observations suggest that the transport properties of such a
polymer network are affected by a competition of two main features. (\emph{i}%
) When the total segment density is fixed, the fraction of \emph{multiply}
occupied bonds increases since random walks frequently ``retrace'' their
last step. As a result, polymerization generates a larger number of
completely \emph{free }($m=0$) bonds which facilitate particle motion. (%
\emph{ii}) As the polymers increase in length, their radius of gyration also
increases, as $\sqrt{\ell }$. The polymers form extended, cigar-shaped
objects which may enclose domains of unoccupied bonds and block particles
very efficiently, especially for $q\rightarrow 0$. As a result, for $q=0$,
the current is controlled by the underlying polymer percolation problem:\ it
is strictly zero when the polymers form a connected cluster of occupied
bonds transverse to the density gradient. Exploring the associated
percolation threshold, $\rho _{c}(\ell )$, as a function of $\ell $, we find
that increasing the fraction of free bonds tends to shift it to higher
densities while creating extended objects and ``loops'' pushes it towards
lower densities; hence, $\rho _{c}(\ell )$ is actually non-monotonic in $%
\ell $. Not surprisingly, the behavior of the currents is very subtle:\ For $%
\rho =0.5$ (well below the percolation threshold), the current decreases
with $\ell $, as larger polymers block more efficiently; for $\rho =0.71$,
the current is non-monotonic, reflecting the behavior of $\rho _{c}(\ell )$.

For nonzero $q$, specifically $q=0.1$, the picture changes significantly.
The finite extent of the polymers becomes far less important when particles
can cross occupied bonds with nonzero probability. As a result, the dominant
factor is now the number of bonds which allow for high particle throughput,
namely, $m=0$ bonds. Since their number increases with $\ell $, the currents
generally increase. Of course, the picture might become more complex again
when the segment density becomes excessively large, generating a large
number of ``highly occupied'' bonds.

To summarize, the transport properties of a random walk network are highly
nontrivial. Clearly, we have explored only a small domain of the full
parameter space, and our statistical errors are still large. Moreover, to
reflect real polymers, more sophisticated models should surely be invoked,
and more surprises may well be in store.

\vspace{0.3cm}

\textbf{Acknowledgements.} It is a pleasure to acknowledge fruitful
discussions with E. Marand and H. Hilhorst. Over the years, both BS and RKPZ
have enjoyed many hours with L. Sch\"{a}fer who taught us much of what we
know about polymers. This work was partially supported by NSF DMR-0088451 
and DMR-0414122. GM thanks the Max Planck Gesellschaft for financial support
and computational resources.

\end{document}